\documentclass[reprint, 
groupedaddress,
superscriptaddress,
longbibliography,
amsmath,amssymb,
aps,
prx,
]{revtex4-2}
\usepackage[colorlinks = true,]{hyperref}
\usepackage{float}
\usepackage{graphicx}
\usepackage{dcolumn}
\usepackage{bm}
\usepackage{xcolor}
\usepackage{braket}
\usepackage{soul}
\usepackage{comment}
\usepackage{amsmath}

\begin{document}

\preprint{APS/123-QED}

\title{Collective Enhancement of Photon Blockade via Two-Photon Interactions}
\author{Lijuan Dong}
\affiliation{Physics Department, Sapienza University, P.le A. Moro 2, 00185 Rome, Italy}
\author{Aanal Jayesh Shah}
\affiliation{Department of Physics and Astronomy, Purdue University, West Lafayette, IN 47906, USA}
%
\author{Peter Kirton}
\affiliation{Department of Physics and SUPA, University of Strathclyde, Glasgow, G4 0NG, United Kingdom}
%
\author{Hadiseh Alaeian}
\affiliation{Elmore Family School of Electrical and Computer Engineering, Purdue University, West Lafayette, IN 47906, USA}
\affiliation{Department of Physics and Astronomy, Purdue University, West Lafayette, IN 47906, USA}

\author{Simone Felicetti}
\email{simone.felicetti@cnr.it}
\affiliation{Institute for Complex Systems, National Research Council (ISC-CNR), Via dei Taurini 19, 00185 Rome, Italy}
\affiliation{Physics Department, Sapienza University, P.le A. Moro 2, 00185 Rome, Italy}

\date{\today}

\begin{abstract}
Analogous to Coulomb blockade for electrons, photon blockade is a key quantum optical effect in which the presence of one photon prevents the transmission of subsequent ones through a nonlinear medium. Beyond its fundamental interest, photon and multi-photon blockade are actively studied as mechanisms for generating technologically-relevant quantum states of light. Although photon blockade typically requires achieving strong light-matter coupling,  increasing the number of atoms fails to enhance antibunching. Here, we analyze the optical transmission properties of a quantum resonator that embeds a two-photon-coupled ensemble of emitters, combining an approximate analytical approach with full quantum numerical simulations.
We show that when light and matter are coupled via a two-photon interaction, both single- and multi-photon blockade can benefit from a \emph{collective enhancement}. 
 We propose different driving schemes in which the second or third-order correlation functions are strongly suppressed with increasing atom number. Differently from established methods, this collective enhancement of non-classical properties occurs with unitary transmission and is ultimately constrained only by decoherence. This demonstrates that collective two-photon couplings are a powerful mechanism for realizing photon blockade even in platforms where \emph{individual} strong coupling is not achievable.
\end{abstract}

\maketitle

\section{Introduction}
Coulomb blockade is a key phenomenon in mesoscopic physics, occurring when electron transport through a nanoscale conductor is strongly affected by inter-particle interactions, so that the presence of a single electron prevents additional ones from tunneling through~\cite{grabert2013single}. Although photons do not interact directly, a closely related phenomenon can arise in nonlinear optical media, where the simultaneous occupation of a system by multiple photons is suppressed. Known as photon blockade (PB), this purely quantum mechanical phenomenon gives rise to a nonlinear optical response and nonclassical photon statistics~\cite{Imamoglu97}.
 The primary figure of merit used to observe PB is the second-order correlation function $g^{(2)}(0)$, which is smaller than one for antibunched light~\cite{Souza2013}. Analogously, multi-PB is characterized by higher-order correlation functions.  
 Currently, intense research efforts are dedicated to reaching PB in new systems and optimizing it in established platforms. Beyond its fundamental interest, PB is a relevant resource for the generation of single- or few-photon states.

Photon blockade can be most directly achieved by coupling an atom or artificial atom to a resonator, producing an anharmonic energy ladder that prevents multiple identical photons from being absorbed~\cite{Imamoglu97}. This method has been experimentally observed in a wide variety of systems, such as atoms in a cavity~\cite{Birnbaum2005}, quantum dots~\cite{Hennessy2007, Muller2015}, and superconducting circuits~\cite{Hoffman2011}.  However, its efficiency is proportional to the light-matter coupling strength of individual quantum emitters, which is bounded by both fundamental and practical reasons.  
 An alternative method, dubbed \emph{unconventional} PB~\cite{Liew2010, Flayac17}, is based on destructive interference between virtual transitions. Validated experimentally~\cite{Radulaski2017, Snijders2018, Vaneph18}, unconventional PB makes it possible to achieve antibunched transmission even in the weak-coupling regime. Different driving schemes have been conceived to induce interference-based PB in systems with Kerr~\cite{Tang2021, mahana2025} and optomechanical~\cite{Komar2013,li2019nonreciprocal} nonlinearities. However, methods based on quantum interference are very sensitive to experimental imperfections, and even slight deviations in detuning, coupling strengths, or loss rates can disrupt the interference and wash out the photon antibunching.  It is well known that increasing the number of emitters does not solve the need for large coupling, as the standard PB is washed away in the many-atom limit~\cite{Carroll2021, Chen2022}. An exception has been identified~\cite{Trivedi2019, Hou23} and observed~\cite{Marinelli2025} for large emitter-resonator detuning, where $g^{(2)}(0)$ can decrease with the atom number $N$ and can be smaller than 1 (antibunched) also in the thermodynamic limit $N\to \infty$. Nevertheless, this phenomenon is still interference-based, which takes place only when the transmission is extremely small (scaling inversely with $N$). Consequently, a stringent trade-off arises between the purity and the brightness of the generated single photons.

A promising direction for enhancing PB lies in the exploration of nonlinear light–matter interactions, which can be realized with superconducting artificial atoms~\cite{felicetti_two-photon_2018, PhysRevA.98.053859, PhysRevLett.121.060503}, quantum dots~\cite{meguebel2025}, and nanomechanical devices~\cite{Zhou2006}, or engineered through parametric pumping and related schemes in solid-state ~\cite{Steele_sidebands, Wang2016, Munoz2018} and atomic~\cite{felicetti_spectral_2015, Schneeweiss_2018, Dareau_2018,casanova2018connecting, PhysRevA.97.023624, PhysRevA.99.032303, Cong_selective} systems. Over the past years, various forms of nonlinear light–matter coupling have been investigated from multiple perspectives. The emergence of quantum phase transitions~\cite{garbe_superradiant_2017, PhysRevA.97.053821, PhysRevA.100.033608, garbe2020dissipation,  Cui2020, li2022nonlinear, shah2024} and collective quantum-emission phenomena~\cite{Delmonte_2ph_battery, Piccione2022} have been investigated, extensively. Their rich phenomenology can be harnessed for quantum-information applications such as quantum gates~\cite{Alushi23}, quantum sensing ~\cite{mutter_transmission-based_2023, ying2025critical}, the generation of frequency-entangled states~\cite{meguebel2025}, GHZ states ~\cite{PhysRevA.99.023854}, cat states~\cite{TPE_Cat}, and Fock states~\cite{PhysRevLett.122.123604}. Strong quadratic emitter–field couplings can give rise to higher-order quantum optical nonlinearities~\cite{felicetti_two-photon_2018, zou2019multiphoton, Wang2021}, hence there has been a growing interest in exploring the PB effect with two-photon coupled models~\cite{Zhang23, Li2024, Liu2024, Solak2025, Zhou25, Zou2020, Liao2013}.
However, all known approaches to PB require either a strong \emph{individual} coupling or a fragile interference between virtual processes.

Here, we show that a \emph{collective} enhancement of both single- and multi-photon blockade can be achieved, using an ensemble of emitters coupled to a single-mode quantum resonator through two-photon interactions. We theoretically analyze the optical transmission properties of the two-photon Tavis-Cummings model, using both analytical and numerical methods. We first use a Holstein-Primakoff approach and a non-Hermitian Hamiltonian approximation to derive analytical solutions for the system's steady state. This analytical approach (labeled NH in the manuscript) is valid under weak drivings and in the thermodynamic limit, i.e., for a large number of emitters $N\to\infty$. Remarkably, we find that the minimum of the second-order correlation function scales as $g^{(2)}(0) \propto 1/N^2$.  Differently from previous studies, this collective enhancement takes place in a resonant, high-transmission regime. 
The analytical treatment clarifies the origin of the collective enhancement of PB, which is not washed out even in the thermodynamic $N\to\infty$ limit, as the nonlinearity comes from the interaction itself.
We compare approximate analytical results with numerical simulations performed on two models of increasing complexity: the Holstein-Primakoff (HP) model, valid in the thermodynamic limit without the weak-drive assumption, and the full-quantum model, valid even for finite-size systems. The results of numerical simulations are in good agreement with analytical solutions, and confirm that the collective advantage is present also for small system sizes. Since the \emph{individual} coupling is not a constraint in the proposed scheme, the main factor limiting the quality of single-PB and multi-PB is atomic dephasing. 
The work is structured as follows. In Sec.~\ref{sec: the model}, we develop a hierarchical modeling approach for an open cavity–emitter ensemble system: with full quantum (FQ) model we refer to exact numerical simulations of the complete master equation, including individual dissipation and dephasing; the Holstein-Primakoff (HP) model is valid in the large $N$ limit, it allows for numerical simulations with much larger systems sizes but it can only include collective decay processes; the non-Hermitian (NH) model is obtained as a further simplification of HP, and it allows for analytical solutions.  In Sec.~\ref{sec:Results}, we compute the transmission and the second-order correlation function using input–output theory, and compare analytical solutions from the NH model with numerical results based on the FQ and HP descriptions. We study both cavity and emitter driving configurations under linear and two-photon coupling, varying the system size, the drive intensity, and the dissipation and decoherence rates. In Sec.~\ref{sec: Conclusion}, we briefly discuss the impact of our analysis and future research directions.


\begin{figure*}[t]
        \centering
        \includegraphics[width=1\linewidth]{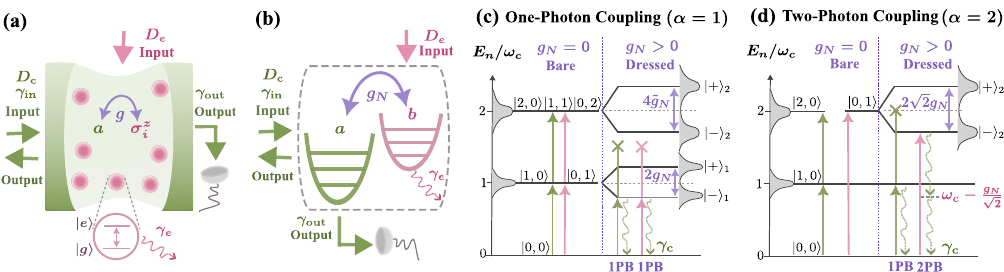}
        \caption{\textbf{Modeling approach and energy-level structure of the coupled cavity--emitter system.}  \textbf{(a)} Physical scheme: A cavity mode \(a\) (green) couples to \(N\) identical two-level emitters \(\sigma_i^z\) (pink) with individual coupling strength \(g\). The cavity interacts with input/output ports at rates \(\gamma_{\mathrm{in}}\) and \(\gamma_{\mathrm{out}}\), and each emitter undergoes individual decay at rate \(\gamma_{e}\).  \textbf{(b)}  
        An effective bosonic model used in the HP approximation. The emitter ensemble is mapped to a collective bosonic mode \(b\)(pink), which interacts with the cavity mode \(a\) via collective coupling \(g_N = g\sqrt{N}\). The effective decay rate of the bosonic mode is \(\gamma_e \).
        \textbf{(c),(d)} Level diagrams for the one-photon (\(\alpha = 1\))and two-photon (\(\alpha = 2\)) coupling cases, respectively. In (c), the first excited doublet \(|\pm\rangle_1\) enables resonant excitation via either cavity or emitter drive when \(\omega_c = \omega_e\), leading to 1PB. In (d), cavity drive still enables 1PB via the \(|\pm\rangle_2\) polariton, while a resonant emitter drive at \(2\omega_c - \sqrt{2}g_N\) induces 2PB with photon-pair emission at \(\omega_c - \frac{g_N}{\sqrt{2}}\). Photons are emitted through the cavity decay channel \(\gamma_c = \gamma_{\mathrm{in}} + \gamma_{\mathrm{out}}\).}

        \label{fig: system-model}
\end{figure*}

\section{The Model}
\label{sec: the model}
This section presents a hierarchical description of the models used here to describe the open cavity–emitter ensemble system illustrated in Fig.~\ref{fig: system-model}. The analysis begins with the full-quantum (FQ) model, which includes individual decay and decoherence processes. A simplified description can be found in the thermodynamic limit, where the Holstein-Primakoff (HP) approximation is applied by mapping collective spin operators onto a single bosonic mode. In the HP model, emitter dissipation is modeled collectively, with individual decay replaced by an effective collective decay channel. Finally, considering the weak-driving limit, we develop a non-Hermitian (NH) formulation where quantum jumps are neglected, which allows us to analytically describe the steady-state transmission properties.

\subsection{Full Quantum Description}
The system consists of a single-mode optical cavity coupled to an ensemble of \( N \) identical two-level quantum emitters (atoms), as shown in Fig.~\ref{fig: system-model}(a). The light–matter interaction is described by the generalized Tavis–Cummings (TC\(_\alpha\)) Hamiltonian, where \( \alpha \) defines the order of the photon exchange process between the cavity and the atomic ensemble. Specifically, \( \alpha = 1 \) corresponds to the standard Tavis–Cummings model with linear coupling, while \( \alpha = 2 \) describes two-photon coupling. 

In the frame rotating at the drive frequency \( \omega_d \), and under the rotating wave approximation (RWA), the coherent dynamics are governed by the driven TC\(_\alpha\) Hamiltonian:
\begin{equation}
\begin{split}
\label{eq:GTC} 
H_{\mathrm{TC}}^\alpha = &\,\Delta_c\, a^\dagger a + \Delta_e \sum_{i=1}^{N} \sigma_i^z \\
&+ g \sum_{i=1}^{N} \left( a^{\dagger \alpha} \sigma_i^- + a^\alpha \sigma_i^+ \right) + H_d,
\end{split}
\end{equation}
where $H_d=D_c(a^\dagger + a)$ for the cavity drive case and $H_d=\sum_{i=1}^{N}\frac{D_e}{\sqrt{N}}(\sigma^+_i+ \sigma^-_i)$ for the collective emitter drive. We define the annihilation (creation) operator of the cavity mode as $a$ ($a^\dagger$), and the Pauli operators for the $i$th emitter as $\sigma_i^z$, $\sigma_i^\pm$, representing population inversion, and raising (lowering) transitions, respectively. The detunings are defined as $\Delta_c = \omega_c - \omega_d$ and $\Delta_e = \omega_e - \alpha \omega_d$, with $\omega_c$, $\omega_e$, and $\omega_d$ denoting the cavity, emitter, and drive frequencies, respectively. Throughout the paper, we assume resonant light–matter interactions, meaning that the emitter frequency is chosen as $\omega_e=\alpha \omega_c $. The individual coupling strength to the cavity is $g$ for all the emitters, the driving amplitude is $D_c$ or $D_e$, and $N$ denotes the number of emitters, also referred to as the system size. We model the open-system dynamics with the Lindblad master equation,
\begin{equation}
\label{ME_FQ}
\dot \rho = -i [H_{\mathrm{TC}}^\alpha, \rho] 
+ \gamma_c\, \mathcal{D}[a](\rho) 
+ \gamma_e \sum_{i=1}^{N} \mathcal{D}[\sigma_i^-](\rho),
\end{equation}
where the Lindblad dissipator is defined as
\(\mathcal{D}[O](\rho) = O \rho O^\dagger - \tfrac{1}{2} (O^\dagger O \rho + \rho O^\dagger O)\).  
We omit \((\rho)\) in the remainder for brevity. The system exhibits two primary loss channels: cavity photon decay at rate $\gamma_c = \gamma_{\mathrm{in}} + \gamma_{\mathrm{out}}$, where $\gamma_{\mathrm{in(out)}}$ are the decay rate from input (output) mirrors, and spontaneous emission from each emitter at rate $\gamma_e$.

This FQ model represents the most general description of the considered system, and it serves as the foundation for both numerical simulation and further analytical treatment. The size of the associated Hilbert space grows exponentially with the number of emitters, scaling as $\mathcal{O}(d_c \times 2^N)$, where $d_c$ is the photon number truncation. In presence of individual decay processes $\mathcal{D}[\sigma_i^-]$, the evolution cannot be reduced to Dicke states. To address this computational challenge, we exploit the permutation symmetry at the density matrix level~\cite{Kirton2017PRL,Kirton2018Superradiant,Shammah2018}, reducing the Hilbert space growth from exponential $O(2^N)$ to $O(N^3)$, without introducing approximations. 
This method allows for an exact treatment of the inclusion of individual decay channels for the emitters, but, unavoidably, the maximum number of emitters that can be included in the simulation is still severely limited by computational resources.

\subsection{Holstein–Primakoff Approximation}

To simplify the FQ model in the thermodynamic limit, we apply the HP transformation~\cite{Holstein1940, Persico1975, Frasca2003}, which maps the collective spin degrees of freedom onto an effective bosonic mode. This model will also be solved numerically. The collective spin operators \( J_\pm = \sum_{i=1}^{N} \sigma_i^\pm \), \( J_z = \frac{1}{2} \sum_{i=1}^{N} \sigma_i^z \), describe an ensemble of \( N \) two-level emitters. In the low-excitation regime, where the number of collective excitations is small compared to the system size, \( \langle b^\dagger b \rangle \ll N \), we adopt the standard spin-down convention and define a bosonic annihilation operator \( b \) that lowers the number of collective spin excitations. The spin operators then map to:
$
J_- \approx \sqrt{N}\, b - \frac{1}{2\sqrt{N}}\, b^\dagger b\, b, \quad 
J_+ \approx \sqrt{N}\, b^\dagger - \frac{1}{2\sqrt{N}}\, b^\dagger b^\dagger b, \quad
J_z = b^\dagger b - \frac{N}{2}. 
$

We explicitly include first-order corrections—nonlinear terms in the excitation number—to account for saturation effects that arise in the finite-size systems, i.e., when $N$ is not asymptotically large. The resulting effective model, illustrated in Fig.~\ref{fig: system-model}(b), consists of two resonators interacting with an effective coupling strength \( g_N \). In the figure, both modes are represented as harmonic oscillators with frequencies \( \omega_c \) and \( \omega_e \).

Substituting these expressions into the TC$_\alpha$ Hamiltonian yields the effective model:
\begin{equation}
\label{eq:H_HPcd}
\begin{split}
H_{\text{HP}}^\alpha = &\Delta_c a^\dagger a + \Delta_e b^\dagger b + g_N (a^{\dagger\alpha} b + a^{\alpha} b^\dagger) \\&- \chi_N \left[(b^\dagger b) a^{\dagger\alpha} b + a^{\alpha} b^\dagger (b^\dagger b)\right] + H_d.
\end{split}
\end{equation}
Under the HP transformation, $H_d$ is not changed for the cavity drive case, while we consider the first order transformation of $H_d= D_e (b^\dagger + b) $ for the collective emitter drive, as we are interested in the weak-drive regime. Here, \( g_N = g\sqrt {N} \) captures the collective enhancement of the light–matter coupling, and \( \chi_N = g / (2\sqrt{N}) \) accounts for nonlinear saturation effects arising at a finite \( N \). While \( g_N \) increases monotonically with the number of emitters, the correction term \( \chi_N \) vanishes in the thermodynamic limit. This implies that the linear-coupling model becomes effectively harmonic at large $N$. However, in the nonlinear-coupling case ($\alpha = 2$), the intrinsic two-photon interaction still induces strong spectral anharmonicity, preserving the system’s nonlinear response even when $\chi_N \!\to\! 0$. The quantitative impact of this first-order correction will be analyzed in the numerical results presented in the following sections. The open-system dynamics of the effective HP model are governed by the following Lindblad master equation,
\begin{equation}
\label{eq:ME_HP}
\dot\rho = -i [H_{\text{HP}}^\alpha, \rho] 
+ \gamma_c\, \mathcal{D}[a] 
+ \gamma_e\, \mathcal{D}[b].
\end{equation}
As individual spin decay processes cannot be included under HP approximation, we include dissipation of the collective spin excitations with an effective rate \( \gamma_e \), as sketched in Fig.~\ref{fig: system-model}(b).

\paragraph*{Spectral structure and photon blockade.}
We now analyze the energy spectrum of the effective Hamiltonian in the absence of drive and dissipation (\(H_d = 0\), \(\gamma_c = \gamma_e = 0\)), focusing on the first two excitation manifolds for both the linear (\(\alpha = 1\)) and nonlinear (\(\alpha = 2\)) coupling cases. The details can be found in Appendix~\ref{app:HP_spectrum}. The total excitation number $ N_\alpha = a^\dagger a + \alpha\, b^\dagger b $ is conserved, and the Hamiltonian becomes block-diagonal in subspaces with fixed excitation number. 

For the linear coupling case, the first excited subspace hybridizes into the polariton doublet: $|\pm\rangle_1 = \frac{1}{\sqrt{2}} \left( |1,0\rangle \pm |0,1\rangle \right),$ with corresponding eigenenergies $E_1^\pm = \omega_c \pm g_N$. The second excited subspace yields two bright states \(|\pm\rangle_2\) with energies \(E_2^\pm = 2 \left(\omega_c \pm \tilde{g}_N \right) \), and one uncoupled dark state. In this subspace the effective polariton splitting is $\tilde{g}_N = \sqrt{(g_N^2 + (g_N - \chi_N)^2)/2}$. As illustrated in Fig.~\ref{fig: system-model}(c), the system can be resonantly excited to the first polariton state $|\pm\rangle_1$ via either cavity or direct emitter drive. Owing to the anharmonicity of the spectrum, quantified by 
$\delta_{\text{anh}} = \tilde{g}_N - g_N$,  transitions to higher excited states become off-resonant, enabling 1PB characterized by antibunched emission at drive frequencies near \((\omega_c \pm g_N)\). This blockade effect is prominent only for moderate values of \(N\), where the system retains significant nonlinearity. In the large-\(N\) limit, the saturation-induced correction vanishes ($\chi_N \rightarrow 0$), so the effective coupling reduces to \(\tilde{g}_N \to g_N\), yielding $\delta_{\text{anh}} \to 0$, and the system spectrum becomes harmonic.

For the two-photon coupling case, the lowest nontrivial manifold is \(n = 2\), which yields the polariton doublet: $
|\pm\rangle_2 = \frac{1}{\sqrt{2}} (|2,0\rangle \pm |0,1\rangle), \quad E_2^\pm = 2\omega_c \pm \sqrt{2}g_N.$
As shown in Fig.~\ref{fig: system-model}(d), the first excited state $|1,0\rangle$ can be selectively populated via cavity drive (green arrow), resulting in 1PB due to the absence of resonant transitions to higher levels (see Sec.~\ref{Cavity_drive}). In this nonlinear coupling regime, the spectrum remains intrinsically anharmonic, with the level spacing deviation quantified by $\delta_{\text{anh}} = g_N$. 
This finite anharmonicity reflects the nonlinear nature of the two-photon interaction rather than emitter saturation and persists even for large $N$.
On the other hand, a collective emitter drive resonantly addresses (pink arrow) the two-excitation polariton states  $|\pm\rangle_2$, resulting in the emission of photon pairs at frequency \((\omega_c \pm \frac{g_N}{\sqrt{2}})\). In section \ref{Emitter_drive}, we will show that in this case, 2PB can be induced.

\subsection{Non-Hermitian Hamiltonian Approximation}
\label{sec:NHmodel}
The HP model provides insight about the spectral structure and, most importantly, allows for numerical simulations in the large $N$ limit. However, it remains analytically intractable. We consider now the weak-driving limit, where the system remains near its vacuum state and excitation events are rare. Under this condition, quantum jumps can be neglected, and the system dynamics can be approximated as \cite{Liew2010}:
\begin{equation}
H_{\text{NH}}^{\alpha} = H_{\text{HP}}^\alpha - \frac{i}{2} \gamma_c a^\dagger a - \frac{i}{2} \gamma_e b^\dagger b.
\label{eq:ME_NH}
\end{equation}
The imaginary terms induce the non-unitary evolution, causing the wavefunction norm to decay over time. This decay reflects that the system remains in a no-jump trajectory throughout its evolution. This approximation preserves the essential spectral structure and nonlinear level spacing responsible for the PB phenomena, while greatly reducing the computational cost. By solving the time-independent Schrödinger equation in Eq.~\eqref{eq:ME_NH}, the NH model can be used to analytically evaluate steady-state observables, such as the transmission spectrum and photon statistics. In the following, we outline the analytical solution procedure; a detailed derivation is presented in Appendix~\ref{app:NH_derivation}. Importantly, this framework also allows one to identify the validity range of the NH model.

To gain analytical insight, we solve the non-Hermitian Schrödinger equation 
\begin{equation}
i\frac{d}{dt}|\psi(t)\rangle = H_{\text{NH}}^\alpha |\psi(t)\rangle.
\label{eq: sc_eq}
\end{equation}
We consider a weak-driving regime and restrict the dynamics to the low-excitation sector of the Hilbert space. In particular, we retain only basis states with total excitation number \(i + \alpha j \leq 2\), leading to the ansatz
\begin{equation}
|\psi(t)\rangle = \sum_{i + \alpha j \leq 2} C_{ij}(t)\, |i,j\rangle.
\label{eq:ansatz_general}
\end{equation}
This ansatz is independent of the drive configuration (cavity or emitter). Inserting it into Eq.~\eqref{eq: sc_eq} and applying the steady-state condition \(\dot{C}_{ij} = 0\), we obtain a closed set of coupled linear equations for the coefficients \(C_{ij}\). Under weak driving, we assume the standard amplitude hierarchy: $|C_{00}| \gg |C_{10}|, |C_{01}| \gg |C_{20}|, |C_{02}|, |C_{11}|$ for $\alpha = 1$, and $|C_{00}| \gg |C_{10}| \gg |C_{20}|, |C_{01}|$ for $\alpha = 2$. Photon correlation functions are computed directly from the steady-state amplitudes: truncation to the two-excitation manifold suffices to capture antibunching in \(g^{(2)}(0)\), while higher-order correlations such as \(g^{(3)}(0)\) require extending the ansatz to include three-excitation states in the emitter-driven two-photon case. Full details of the ansatz and resulting equations are given in Appendix~\ref{app:NH_derivation}.

Let us begin by commenting on the validity of this approach. First, when the drive is strong, the number of excitations grows and quantum jumps become non-negligible, leading to deviations from the NH predictions. Second, even under the weak-drive assumption, the validity of the NH approximation must be carefully examined in regimes where quantum jumps significantly contribute to the system’s excitation pathways. For instance, in the case of emitter driving with nonlinear two-photon coupling, the NH method yields \( C_{10} = 0 \), in contradiction to the full master-equation results. This discrepancy arises because the state \( |10\rangle \) is not populated through coherent dynamics but is instead generated via dissipation (quantum jumps).

To restore physical consistency, we apply an excitation-space normalization condition of the form
$
1 - |C_{00}|^2 \approx |C_{10}|^2 + |C_{20}|^2 + |C_{01}|^2 \sim D_e^2,
$
which effectively accounts for dissipative contributions to the steady-state solution. A detailed discussion of this renormalization and the criteria for using the NH approximation appears in Appendix~\ref{app:B}.
 This refinement improves the agreement with numerical simulations and helps delineate the applicability of the NH model. In the following section, we proceed to analyze the steady-state observables derived from the NH approximation, comparing them with numerical results across different drive configurations, drive strengths, coupling types, and dissipation channels.

\begin{figure*}[!]
        \centering   \includegraphics[width=\textwidth]{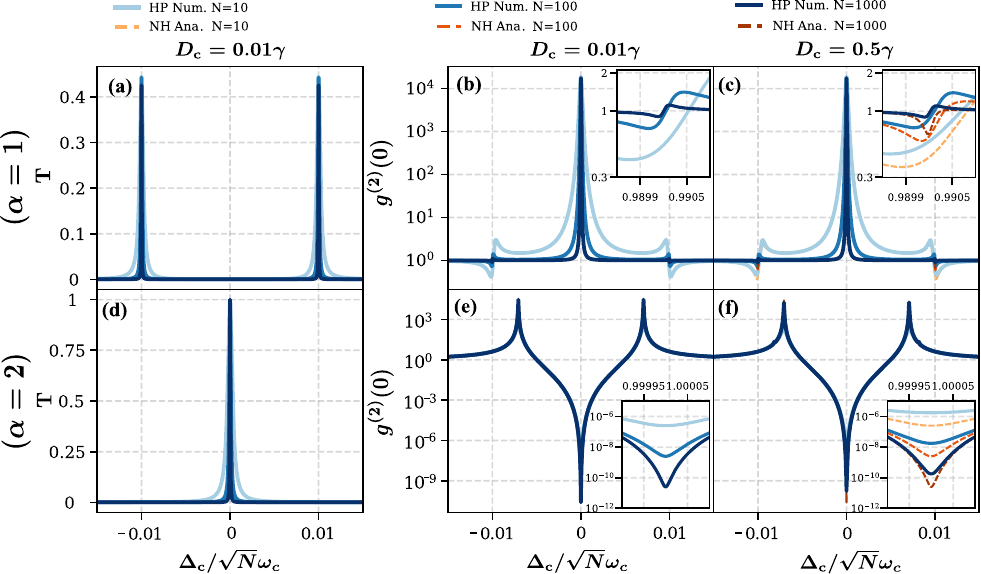}
        \caption{\textbf{One- and two-photon TC ($\alpha=1, 2$). Comparison of NH analytical and HP numerical spectra under coherent cavity drive.} One-photon coupling (\(\alpha = 1\)) and two-photon coupling (\(\alpha = 2\)) are shown in the first and second rows, respectively. For each case, the normalized transmitted intensity \(T\) and second-order correlation function \(g^{(2)}(0)\) are plotted as functions of detuning  \(\Delta_c\) for varying drive amplitudes \(D_c\). Solid lines (HP Num.) show numerical results from the HP model, while dashed lines (NH Ana.) correspond to analytical results from the NH model. Color shading indicates emitter number: \(N = 10\) (light), \(N = 100\) (medium), and \(N = 1000\) (dark). Frequencies are rescaled by \(\sqrt{N}\) to account for collective enhancement. Insets show zoomed-in plots near the polariton resonance. Analytical curves are overlaid but become distinguishable only where deviations from HP numerical results are significant. Parameters: \(D_c = 0.01\gamma\) for (a), (b), (d), (e), and \(D_c = 0.1\gamma\) for (c), (f); \( g= 0.01\,\omega_c \), \(\gamma_c = 2\gamma\), \(\gamma_e = \gamma\), with \(\gamma = 0.001\,\omega_c\).  }
        \label{fig:HP&Nan_C_drive}
\end{figure*}

\section{Results}
\label{sec:Results}
To characterize the output field, we apply the input-output formalism~\cite{walls2008quantum}, assuming weak and frequency-independent coupling to a Markovian environment. We focus on the steady state properties, achieved for a long evolution time, and define $\langle a_{\text{out}}\rangle^{\text{ss}} = \lim_{t \to \infty} \langle a_{\text{out}}(t) \rangle$. The normalized transmission and equal-time photon correlation functions are defined as
\begin{equation}
T = \frac{n_{\text{out}}}{ n_{\text{in}}}, \quad
g^{(n)}(0) = \frac{\langle (a_{\text{out}}^\dagger)^n (a_{\text{out}})^n \rangle^{\text{ss}}}{n_{\text{out}}^n}.
\label{eq: correlation function}
\end{equation}
We define the output photon flux as $n_{\text{out}} = \gamma_{\text{out}} \langle a^\dagger a \rangle^{\text{ss}}$. The input photon flux depends on the driving configuration:  $n_{\text{in}} = D_c^2 / \gamma_{\text{c}}$ for the cavity drive case; $n_{\text{in}} = D_e^2 / \gamma_e$ for the emitter drive case. 

We analyze the intensity and statistics of the transmitted field and then explore how PB phenomena emerge by tuning the drive frequency and strength, the emitter number, and the coupling order $\alpha$. Comparing analytical predictions from the NH model with numerical results from the FQ and HP models enables us to assess the accuracy and limitations of the approximations.

\begin{figure}[t]
        \centering
        \includegraphics[width=1\linewidth]{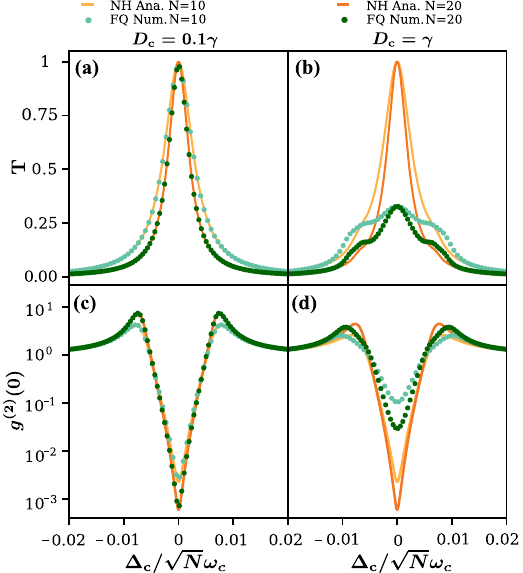}
        \caption{\textbf{Two-photon TC ($\alpha=2$). Comparison of FQ numerical simulations with NH analytical results.} Panels (a), (b) show the transmitted intensity \(T\), while (c), (d) present the second-order correlation function \(g^{(2)}(0)\), under weak and strong cavity driving. NH analytical predictions are shown in orange (light for \(N = 10\), dark for \(N = 20\)); FQ numerical results are shown in green (light for \(N = 10\), dark for \(N = 20\)).  Parameters: Panels (a), (c): weak drive \(D_c = 0.1\gamma\); panels (b), (d): strong drive \(D_c = \gamma\);  \(g= 0.01\,\omega_c\), \(\gamma_c = \gamma_e = \gamma = 0.01\,\omega_c\),\(\gamma = 0.01\,\omega_c\), same values for Fig.~\ref{fig4: g2-min-vs-N} and Fig.~\ref{fig:dep}. }
     
        \label{fig3:FQ vs NH}
\end{figure}

\subsection{Cavity drive}
\label{Cavity_drive}
We begin by analyzing the response of the system under coherent cavity drive. Figure~\ref{fig:HP&Nan_C_drive} shows the transmission, $T$, and second-order correlation function, $g^{(2)}(0)$, vs.~the drive detuning, $\Delta_c$, for both linear ($\alpha=1$) and nonlinear ($\alpha=2$) coupling, in the top and bottom rows, respectively.  Results from the HP model (solid lines) are compared with analytical solutions from the NH approximation (dashed lines), for various numbers of emitters $N$ and drive strengths $D_c$.
The transmission features observed in Fig.~\ref{fig:HP&Nan_C_drive}(a),(d) can be understood by comparing the system’s energy spectra shown in Figs.~\ref{fig: system-model}(c),(d). For $\alpha = 1$, the HP numerical results show a vacuum Rabi splitting, with two symmetric peaks around $\Delta_c= \pm g_N$. This splitting arises from the hybridization between the cavity and the collective atomic mode, which produces dressed polaritonic doublets separated by $2g_N$. The peak structure in the weak drive regime directly reflects the allowed transitions into these lowest-energy doublets. In contrast, for $\alpha = 2$, the transmission exhibits a single peak centered at resonance, with no visible splitting at weak drive. This behavior stems from the fact that the first excited state $|1,0\rangle$ does not involve atomic excitation, and so it is decoupled from the emitter ensemble. As a result, the transmission intensity resembles that of an empty cavity, despite the presence of the two-photon coupling. In both cases, the HP numerical spectra show excellent agreement with the NH analytical results. As detailed in Appendix~\ref{app:NH_derivation}, the NH expression for $T$—particularly for $\alpha = 2$—is dominated by the denominator $(\Delta_c - i\gamma)^2$, yielding a Lorentzian peak (we have set $\gamma_{\mathrm{in}} = \gamma_{\mathrm{out}}= \gamma$,  $\gamma_c=  2\gamma$ and $\gamma_e = \gamma$). For all the considered emitter numbers $N$, the analytical results are in excellent agreement with numerical simulations, confirming the validity of the approximation. 

Let us now focus on the statistics of the emitted field. For linear coupling, Fig.~\ref{fig:HP&Nan_C_drive}(b) shows a dip in $g^{(2)}(0)$ near the polariton resonances at detunings $\Delta_c = \pm g_N$, indicating single-photon antibunching. However, this effect becomes less pronounced as the emitter number $N$ increases (shown in the inset). This trend arises because the effective Kerr nonlinearity $\chi_N$—responsible for the energy-level anharmonicity—diminishes with increasing $N$, even though the collective coupling $g_N$ becomes stronger. From a physical perspective, this reflects the approach to the bosonic limit, where the collective spin excitation becomes increasingly harmonic and PB fades away~\cite{Trivedi2019}. 

\begin{figure}[h]
        \centering
        \includegraphics[width=1\linewidth]{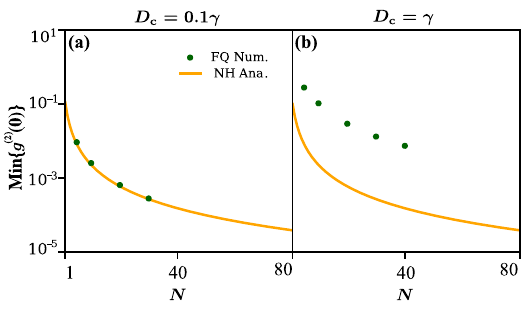}
        \caption{\textbf{Two-photon TC ($\alpha=2$). Minimum of \(g^{(2)}(0)\) for varying emitter number \(N\).} Minimum values of the second-order correlation function \(g^{(2)}(0)\) as a function of emitter number \(N\), for (a) weak drive \(D_c = 0.1\gamma\) and (b) strong drive \(D_c = \gamma\).}

           \label{fig4: g2-min-vs-N}
\end{figure}
\begin{figure}[t]
        \centering
        \includegraphics[width=1\linewidth]{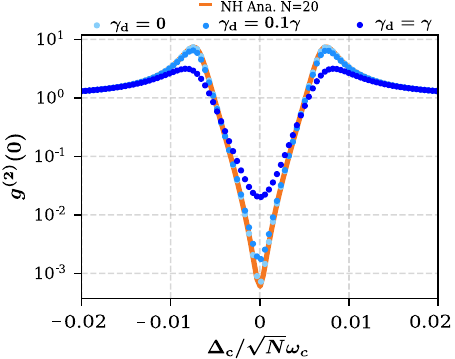}
        \caption{\textbf{Two-photon TC ($\alpha=2$). Effect of emitter dephasing on NH analytical results.} Second-order correlation function \(g^{(2)}(0)\) obtained from NH analytical results (orange solid lines) and FQ simulations (dots), for \(N = 20\) emitters. Parameters: \(D_c = 0.1\gamma\); emitter dephasing rates \(\gamma_d = 0\), \(0.1\gamma\), and \(\gamma\).}
        \label{fig:dep}
\end{figure}
\begin{figure*}[t]
        \centering
        \includegraphics[width=\textwidth]{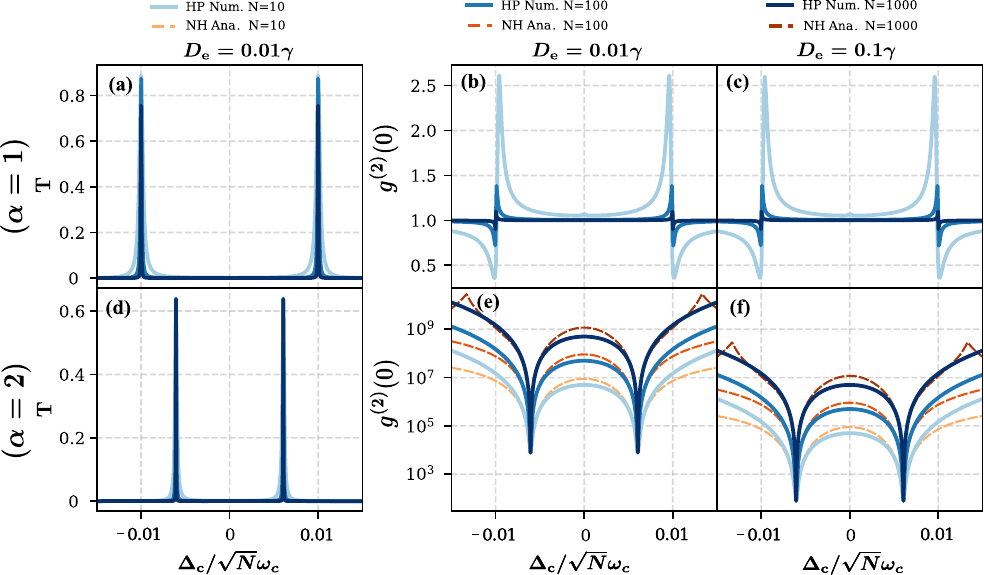}
       \caption{\textbf{One- and two-photon TC ($\alpha=1, 2$). Comparison of NH analytical and HP numerical output spectra with coherent emitter drive.} Same format and parameters as in the cavity-driven case (see Fig. \ref{fig:HP&Nan_C_drive}), except that the figures are plotted as functions of detuning  \(\Delta_c\), meanwhile the drive is applied to the emitter ensemble and the cavity is coupled to a single output port with decay rate $\gamma_c = 2\gamma$.}
        \label{fig:HP&Nan_Q_drive}
\end{figure*}  
We move our focus now to the two-photon coupling case, shown in Fig.~\ref{fig:HP&Nan_C_drive}(e).
In contrast to what is observed for the linear case, for $\alpha=2$ at resonance, $g^{(2)}(0)$ reduces as $N$ increases (visible in the inset). The HP numerical data also confirm this collective enhancement of PB. This behavior is captured quantitatively by the NH analytical expression, which predicts $g^{(2)}(0)=\frac{\gamma^4/4}{g^4 N^2}  \propto 1/N^2$ in the large $N$ limit (see Appendix~\ref{app:NH_derivation}). Thus, it vanishes asymptotically as $N \rightarrow \infty$, indicating near-perfect antibunching in the thermodynamic limit. Finally, in Fig.~\ref{fig:HP&Nan_C_drive}(c) and (f), we consider larger driving strengths. As expected, the NH model becomes less accurate in this regime, since it is valid only for low excitation numbers.

We now benchmark the results obtained for $\alpha=2$ with numerical simulations of the FQ model master equation~\eqref{ME_FQ}, which included individual spin decay. Figure~\ref{fig3:FQ vs NH} compares the FQ and NH predictions of the transmission intensity and photon statistics. We consider a higher dissipation rate ($\gamma = 0.01\omega_c$) to reduce the computational cost. The plots display $T$ and $g^{(2)}(0)$ for $N = 10$ and $N = 20$, under weak ($D_c = 0.1\gamma$) and strong ($D_c = \gamma$) cavity driving. For weak driving, the NH approximation matches well with the FQ simulations, capturing both the resonance peak in transmission and the antibunching dip in the correlation function, as shown in Fig.~\ref{fig3:FQ vs NH}(a) and (c), respectively. Although the minimum $g^{(2)}(0)$ values are higher than in Fig.~\ref{fig:HP&Nan_C_drive} due to increased dissipation, the expected suppression with increasing $N$ remains evident. In summary, comparison with the FQ model shows that the results obtained with the HP and NH models are reliable even for a relatively small number of emitters and for moderate driving strength. However, for strong driving Fig.~\ref{fig3:FQ vs NH}(b) and (d)—when the driving strength approaches the cavity decay rate— the low-excitation assumption underlying the NH approximation breaks down, leading to deviations from the FQ results.
  
To further analyze the collective enhancement of the PB in the nonlinear coupling case, Figure~\ref{fig4: g2-min-vs-N} presents the scaling of the minimum $g^{(2)}(0)$ at resonance as a function of emitter number $N$, under both weak and strong coherent cavity driving. In the weak-drive regime Fig.~\ref{fig4: g2-min-vs-N}(a), NH and FQ results show excellent agreement, confirming the $1/N^2$ dependence of the $g^{(2)}(0)$  predicted analytically.  At stronger driving Fig.~\ref{fig4: g2-min-vs-N}(b), the NH model overestimates the blockade effect and its predictions deviate from the FQ results. Nevertheless, the FQ data clearly confirm that the collective enhancement of PB persists with increasing $N$.

Finally, we analyze the impact of dephasing. Figure~\ref{fig:dep} shows the $g^{(2)}(0)$ evaluated numerically using the FQ model, for different values of the individual dephasing rate. The analytical result with the NH model without dephasing is left as a reference. For moderate values of $\gamma_d$ the minimum of the $g^{(2)}(0)$ is only slightly lifted. A dephasing comparable to the decay rate has a relevant impact, but the value of $g^{(2)}(0)$ is still significantly below unity. This confirms that PB remains robust even in the presence of moderate decoherence. In principle, the collective enhancement of PB via two-photon coupling can be made arbitrarily strong by increasing the number of emitters. However, this scaling will eventually be limited by decoherence rates. Accordingly, when two-photon couplings are considered, PB can in principle be arbitrarily enhanced by increasing the number of emitters. For a fixed $N$, the relevant limiting factor will be given by the dechorence rates.


\subsection{Emitter drive}
\label{Emitter_drive}
 
\begin{figure}[t]
        \centering
        \includegraphics[width=1\linewidth]{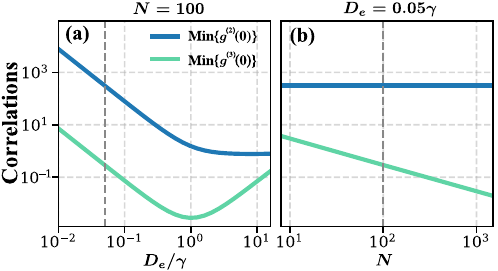}
        \caption{\textbf{Two-photon TC ($\alpha=2$). Photon correlation functions in the two-photon HP model under emitter drive.} 
        \textbf{(a)} Correlation functions \(g^{(2)}(0)\) and \(g^{(3)}(0)\) as functions of drive amplitude \(D_e/\gamma\) for \(N = 100\). 
       \textbf{(b)} Same quantities as functions of emitter number \(N\), with fixed drive amplitude \(D_e = 0.05\,\gamma\). 
       Drive frequency is set to the lower polariton resonance with \(\Delta_c =g_N/\sqrt{2}\). 
       Parameters: \(g = 0.01\omega_c\), \(\gamma_c = 2\gamma\), \(\gamma_e = \gamma\), \(\gamma = 0.001\,\omega_c\).}
       \label{fig:g2g3_HP}
\end{figure}  

To complete the picture, in this sub-section, we compare the linear ($\alpha=1$) and nonlinear ($\alpha=2$) couplings with the drive applied to the emitters instead of the cavity. We use the HP model of  Eq.~\eqref{eq:H_HPcd} with the emitter drive included. The resulting transmission spectra are shown in Fig.~\ref{fig:HP&Nan_Q_drive} (a) and (d). 
Notice that driving the emitters entails a frequency-conversion process for $\alpha=2$, so the transmission coefficient must be renormalized as $T = n_{\text{out}}/\alpha n_{\text{in}}$.

Unlike in the cavity-driven case, when the drive is applied to the emitters the transmission spectra exhibit a characteristic double-peak structure even for two-photon coupling. This is a result of selection rules, which link the ground state to the polaritonic doublet obtained by hybridizing the states $\ket{2,0}$ and $\ket{0,1}$, as shown in Fig.~\ref{fig: system-model}(d). The HP numerical results shown in Fig.~\ref{fig:HP&Nan_Q_drive} agree well with the analytical predictions, with overlapping spectral features and lineshapes observed under weak driving. 

A clear distinction between the one- and two-photon HP models emerges under emitter drive in the photon statistics of the emission field. In the one-photon coupling case, \( g^{(2)}(0) \) falls below unity near each polaritonic resonance, indicating sub-Poissonian statistics and the suppression of multiple excitations, we observe the standard properties of one photon blockade that, as expected, is stronger for smaller $N$ and gradually fades away as the ensemble size increases. In contrast, the two-photon coupling model exhibits super-Poissonian statistics across the spectrum, with \( g^{(2)}(0) > 1 \), reflecting a higher probability of multi-photon events. As shown in the following, the system exhibits signatures consistent with a two-photon blockade — correlated emission of photon pairs accompanied by suppression of higher-order processes.

To characterize the emergence of the 2PB regime in the emitter-driven case with nonlinear coupling, we analyze the third-order correlation function \( g^{(3)}(0) \) alongside \( g^{(2)}(0) \). The 2PB regime is identified by \( g^{(2)}(0) \gtrsim 1 \) and \( g^{(3)}(0) \ll 1 \), indicating photon-pair bunching with strong suppression of three-photon events. These features are most prominent in the low-excitation limit \( \langle a^\dagger a \rangle \ll 1 \), where quantum nonlinearities dominate and our approximations (e.g., HP and NH) are most reliable. Figure~\ref{fig:g2g3_HP}(a) confirms the presence of 2PB, by showing that a broad region of driving intensity exists where \( g^{(2)}(0) \) remains above unity while \( g^{(3)}(0) \) stays well below one. In Fig.~\ref{fig:g2g3_HP}(b), we fix the emitter drive strength \( D_e = 0.05\,\gamma \) and vary the emitter number \( N \). We find that while \( g^{(2)}(0) \) remains nearly unchanged, \( g^{(3)}(0) \) decreases significantly with increasing \( N \), confirming that also 2PB profits from a collective enhancement of the coupling strength.

Interestingly, Fig.~\ref{fig:g2g3_HP}(a) reveals a pronounced dip in \( g^{(3)}(0) \) as the drive intensity increases, signaling an optimal regime for 2PB. This non-monotonic behavior arises from a competition between two effects: on the one hand, the drive must be strong enough to populate the two-excitation manifold and enable correlated photon-pair emission; on the other hand, overly strong driving can overcome the system’s nonlinearity and populate higher-excitation manifolds, thereby degrading the blockade. As a result, an intermediate drive strength optimally supports 2PB, characterized by enhanced two-photon bunching and suppression of three-photon events. While the position of the dip in \( g^{(3)}(0) \) remains largely insensitive to the emitter number \( N \)—due to the collective drive configuration and intensity rescaling in HP model—the depth of the dip becomes more pronounced with increasing \( N \), reflecting the enhancement of collective nonlinearity. This highlights how strong photon-pair correlations emerge from a subtle interplay between coherent excitation and nonlinear saturation effects in the low-excitation regime.


\section{Conclusion}
\label{sec: Conclusion}
We have identified a many-body cavity-QED system where photon and multi-photon blockade profit from a \emph{collective} enhancement of the light-matter interaction strength. Unlike previous approaches, in this case photon blockade occurs under conditions of high transmission, without a strict trade-off between blockade and visibility. 
This is particularly relevant for all cavity QED systems where including many emitters in an optical resonator is straightforward, while enhancing the \emph{individual} light-matter coupling remains challenging. For example, this is the case for organic~\cite{Khazanov2023Embrace} and inorganic~\cite{QFluids,Delteil2019Towards} exciton polaritons, where the collective light-matter interactions can be pushed even into the ultrastrong coupling regime, but only in the regime where the number of emitters is so large that the nonlinearity of matter excitations is washed out. Realizing two-photon coupling is undoubtedly challenging, but it holds the potential to enable the observation of strong quantum-optical nonlinear effects in atomic, molecular, and solid-state platforms that have so far been restricted to the linear regime. Accordingly, this work opens several promising directions for future basic and applied research. The considered method could be further optimized by incorporating interference effects within a hybrid conventional–unconventional approach. Another avenue for future fundamental research is to study the breakdown of photon blockade, which has been thoroughly explored theoretically~\cite{Carmichael2015,Curtis2022,LeBoite2016FatePhotonBlockade} and experimentally~\cite{Fink2017PhotonBlockadeBreakdown}, though only in systems exhibiting linear interactions. Finally, these results can be applied in the design of new devices that harness collective light–matter interactions to generate nonclassical light for quantum sensing, communication, and computing tasks.

\begin{acknowledgments}
L.D. acknowledges support from the China Scholarship Council (CSC, Grant No. 202106890034). PK acknowledges support from EPSRC (Grant
No. EP/Z533713/1).
Research supported as part of QuPIDC, an Energy Frontier Research Center, funded by the US Department of Energy (DOE), Office of Science, Basic Energy Sciences (BES), under award number DE-SC0025620 (AJS and HA).
In addition, part of the research was supported by the US Department of Energy (DOE), the Office of Basic Energy Sciences (BES), and the Division of Materials Sciences and Engineering under award number DE-SC0025554 (AJS and HA). S.F. acknowledges financial support from the foundation Compagnia di San Paolo, grant vEIcolo no. 121319, and from PNRR MUR project PE0000023-NQSTI financed by the European Union–Next Generation EU. 
\end{acknowledgments}

\clearpage
\newpage
\appendix

\appendix

\section{Spectral Structure of HP model}
\label{app:HP_spectrum}

In this appendix, we analyze the excitation spectrum of the HP model in the absence of driving and dissipation, focusing on the one-photon ($\alpha = 1$) and two-photon ($\alpha = 2$) coupling cases. We examine the structure of the first and second excited manifolds, the role of spectral anharmonicity, and the formation of bright and dark eigenstates. Without external drive or dissipation (\(H_d = 0\), \(\gamma_c = \gamma_e = 0\)), the effective Hamiltonian is 
\begin{equation}
\begin{split}
H_{\mathrm{HP}}^\alpha = & \omega_c a^\dagger a + \omega_e b^\dagger b + g_N (a^{\dagger\alpha} b + a^{\alpha}b^\dagger) \\&- \chi_N \left[(b^\dagger b) a^{\dagger\alpha} b + a^{\alpha} b^\dagger (b^\dagger b)\right],
\end{split}
\end{equation}
which conserves the total excitation number $
N_\alpha = a^\dagger a + \alpha b^\dagger b.
$
This conservation law allows the Hamiltonian to be block-diagonalized into invariant subspaces labeled by \(n = N_\alpha\), each spanned by the Fock product states \(|n_a, n_b\rangle\) satisfying \(n_a + \alpha n_b = n\).
Throughout the paper, we assume resonant light–matter interactions, meaning that the emitter frequency is chosen as $\omega_e=\alpha \omega_c $.
Meanwhile, we define the  relative anharmonicity for the first two-excitation subspace by
\begin{equation}
\delta_{\text{anh}} =
|E_{2}^{\pm} - 2E_{1}^{\pm}|,
\end{equation}
with $ E_0=0 $ to quantify the deviation of the excitation spectrum from a harmonic ladder , which is crucial for characterizing photon blockade and nonlinear optical response.

\subsection{One-Photon Coupling ($\alpha = 1$)}

\paragraph{One-excitation subspace --}
The $n=1$ subspace is spanned by $\{|1,0\rangle$, $|0,1\rangle\}$. The Hamiltonian in this basis is:
\begin{equation}
H^{(1)} =
\begin{pmatrix}
\omega_c & g_N \\
g_N & \omega_c
\end{pmatrix},
\end{equation}
which yields eigenstates:
\begin{align}
|\pm\rangle_1 &= \frac{1}{\sqrt{2}} (|1,0\rangle \pm |0,1\rangle), \\
E_1^\pm &= \omega_c \pm g_N , 
\end{align}
These are the familiar lower and upper polariton states. The energy splitting is $2g_N$, corresponding to the Rabi splitting in the single-excitation manifold.
\\
\paragraph{Two-excitation subspace --}
The $n=2$ subspace includes three basis states: $\{|2,0\rangle$, $|1,1\rangle$, $|0,2\rangle\}$. Including the saturation correction $\chi_N$, the Hamiltonian becomes:
\begin{equation}
H^{(2)} =
\begin{pmatrix}
2\omega_c & \sqrt{2}g_N & 0 \\
\sqrt{2}g_N & 2\omega_c & \sqrt{2}(g_N - \chi_N) \\
0 & \sqrt{2}(g_N - \chi_N) & 2\omega_c
\end{pmatrix}.
\end{equation}
Diagonalizing $H^{(2)}$ yields three eigenstates: two bright polaritons, $|\pm\rangle_2$, and one dark state $|D\rangle$, which remains decoupled from the cavity mode. 
The bright and dark eigenstates are given by
\begin{align}
|+\rangle_2 &= \frac{1}{2} \left( |2,0\rangle + \sqrt{2}\,|1,1\rangle + |0,2\rangle \right), \\
|-\rangle_2 &= \frac{1}{2} \left( |2,0\rangle - \sqrt{2}\,|1,1\rangle + |0,2\rangle \right)\\
|D\rangle &= \frac{1}{\sqrt{2}} (|2,0\rangle - |0,2\rangle),
\end{align}
with corresponding eigenenergies
\begin{equation}
E_2^{\pm} = 2 \left(\omega_c \pm \tilde{g}_N \right), E_D = 2\omega_c,
\end{equation}
where the effective polariton splitting is defined as
\begin{equation}
\tilde{g}_N = \sqrt{(g_N^2 + (g_N - \chi_N)^2)/{2}}.
\end{equation}
The anharmonicity of
the spectrum is $\delta_{\text{anh}} = \tilde{g}_N-g_N$. In the large-\(N\) limit, where the saturation-induced nonlinearity vanishes (\(\chi_N \to 0\)), the effective coupling in the second excitation manifold approaches \(\tilde{g}_N \to g_N\), leading to $\delta_{\text{anh}} = 0$. The collective spin maps onto a bosonic mode, and the model reduces to two linearly coupled oscillators, therefore describes a formally harmonic system.

\subsection{Two-Photon Coupling ($\alpha = 2$)}

\paragraph{Two-excitation subspace --}
The two-photon interaction does not couple the states 
\(|1,0\rangle\) and \(|0,1\rangle\). We therefore focus on the lowest nontrivial excitation manifold of \(n = 2\), which is spanned by the basis states \(\{|2,0\rangle, |0,1\rangle\}\). In this subspace and with $\omega_e=\alpha \omega_c $, the effective Hamiltonian takes the form
\begin{equation}
H^{(2)} =
\begin{pmatrix}
2\omega_c & \sqrt{2}g_N \\
\sqrt{2}g_N & 2\omega_c
\end{pmatrix},
\end{equation}
with eigenstates:
\begin{align}
|\pm\rangle_2 &= \frac{1}{\sqrt{2}} (|2,0\rangle \pm |0,1\rangle), \\
E_2^\pm &= 2\omega_c \pm \sqrt{2}g_N.
\end{align}

Unlike the $\alpha = 1$ case, in the nonlinear two-photon coupling ($\alpha = 2$), the polariton splitting is independent of $\chi_N$ and remains intrinsically anharmonic even in the large-$N$ limit. 
Here, the anharmonicity, $\delta_{\text{anh}} = \sqrt{2}g_N$, originates from the nonlinear light–matter interaction itself rather than from emitter saturation. Consequently, the nonlinearity is collectively enhanced through $g_N = g\sqrt{N}$.

\section{Analytical Derivation of Non-Hermitian Steady States}
\label{app:NH_derivation}

To analytically derive steady-state observables within the HP framework, presented in Secs.~\ref{sec:NHmodel}–\ref{sec:Results}, we solve the time-independent non-Hermitian Schrödinger equation under weak driving. This approach simplifies the dynamics by neglecting quantum jumps, as justified in the Sec.~\ref{sec:NHmodel}.

The total non-Hermitian Hamiltonian includes effective loss terms on both bosonic modes. For both drive types, the effective Hamiltonian is in Eq.~\eqref{eq:ME_NH}. 
We define the complex detunings
\(\tilde{\Delta}_{c(e)}=\Delta_{c(e)}-i\gamma_{c(e)}/2\),

In both the cavity and emitter drive configurations, we compute the steady-state wavefunction by solving the non-Hermitian Schrödinger Eq.~\eqref{eq: sc_eq}. We adopt different low-excitation ansätze in Eq.~\eqref{eq:ansatz_general} depending on the order of light–matter coupling \( \alpha \), and considering Eq.~\eqref{ansatz_3ph} for the third correlation function for the emitter drive case.

The resulting linear systems for the amplitudes \( C_{ij} \) are solved analytically. For brevity, we henceforth suppress the explicit time argument and write $C_{ij}\equiv C_{ij}(t)$. After solving for stationary solutions, using the definitions of the transmission and equal-time second-order correlation in Sec.~\ref{sec:Results} and Eq.~\eqref{eq: correlation function},  we get:

For $\alpha=1$,
\begin{equation}
\label{eq:T_general}
T = \frac{\gamma_{\mathrm{in}} \gamma_{\mathrm{out}}}{ D_c^{\,2}} 
(|C_{10}|^2 +|C_{11}|^2 + 2 |C_{20}|^2),
\end{equation}
\begin{equation}
\label{eq:g2_general}
g^{(2)}(0) = \frac{2|C_{20}|^2}{(|C_{10}|^2 +|C_{11}|^2 + 2 |C_{20}|^2)^2}.
\end{equation}

For $\alpha=2$,
\begin{equation}
\label{eq:T_general2}
T = \frac{\gamma_{\mathrm{in}} \gamma_{\mathrm{out}}}{ D_c^{\,2}} 
(|C_{10}|^2  + 2 |C_{20}|^2),
\end{equation}
\begin{equation}
\label{eq:g2_general2}
g^{(2)}(0) = \frac{2|C_{20}|^2}{(|C_{10}|^2 + 2 |C_{20}|^2)^2}.
\end{equation}

For cavity drive, we take \( D_c \) as the drive amplitude, with \( \gamma_{\mathrm{in}} = \gamma_c \) and \( \gamma_{\mathrm{out}} = \gamma_c \), corresponding to both input and output coupling through the cavity, as illustrated in Fig.~\ref{fig: system-model}. For emitter drive, we replace \( D_c \to D_e \), and assign \( \gamma_{\mathrm{in}} = \gamma_e \) and \( \gamma_{\mathrm{out}} = \gamma_c \), indicating that the input couples through the emitter ensemble while the output is collected from the cavity. In both configurations, we assume \( \gamma_c/2 = \gamma_e = \gamma \).

\subsection{Cavity drive}
\subsubsection{Linear coupling (\(\alpha=1\))}

Inserting the ansatz of Eq.~\eqref{eq:ansatz_general} into the non-Hermitian Schrödinger equation with cavity drive of Eq.~\eqref{eq:ME_NH}, 
We obtain the following set of coupled linear equations for the amplitudes \(\{C_{ij}\}\):
\begin{subequations}\label{eq:cd_lin_eom}
\begin{align}
i \dot{C}_{00} &= D_c\, C_{10}, \label{eq:cd_lin_eom_a}\\
i \dot{C}_{10} &= D_c\, C_{00} + \tilde{\Delta}_c\, C_{10} + g_N\, C_{01} + \sqrt{2}\, D_c\, C_{20}, \label{eq:cd_lin_eom_b}\\
i \dot{C}_{01} &= \tilde{\Delta}_e\, C_{01} + g_N\, C_{10} + D_c\, C_{11}, \label{eq:cd_lin_eom_c}\\
i \dot{C}_{20} &= 2\tilde{\Delta}_c\, C_{20} + \sqrt{2}\, g_N\, C_{11} + \sqrt{2}\, D_c\, C_{10}, \label{eq:cd_lin_eom_d}\\
i \dot{C}_{02} &= 2\tilde{\Delta}_e\, C_{02} + \sqrt{2}\, g_N\, C_{11} - \sqrt{2}\, \chi_N\, C_{11}, \label{eq:cd_lin_eom_e}\\
i \dot{C}_{11} &= D_c\, C_{01} + (\tilde{\Delta}_c+\tilde{\Delta}_e)\, C_{11} \nonumber\\
&\quad + \sqrt{2}\,(g_N-\chi_N)\, C_{02} + \sqrt{2}\, g_N\, C_{20}. \label{eq:cd_lin_eom_f}
\end{align}
\end{subequations}

Under weak pumping \(\bigl(|C_{00}|\gg |C_{10}|,|C_{01}|\gg |C_{20}|,|C_{02}|,|C_{11}|\bigr)\), hence the contribution of the two-excitation amplitudes to the steady-state values of \(C_{20},C_{02},C_{11}\) is negligible. Setting \(\dot{C}_i = 0\), we solve for the steady-state amplitudes in terms of \(C_{00}\),

\begin{align}
\label{eq:cd_lin_C01}
C_{01} &\approx -\frac{g_N}{\tilde{\Delta}_e}\, C_{10},\\
\label{eq:cd_lin_C10}
C_{10} &\approx -\,\frac{D_c}{\tilde{\Delta}_c-\dfrac{g_N^2}{\tilde{\Delta}_e}}\; C_{00}\,.
\end{align}
To leading non-vanishing order in \(D_c\), the two-excitation amplitudes read as:
\newcommand{\Dc}{\tilde{\Delta}_c}
\newcommand{\De}{\tilde{\Delta}_e}
\newcommand{\rt}{\sqrt{2}}

\begin{subequations}\label{eq:twoexc_amps}
\begin{align}
C_{20} =& \frac{D_c}{\rt\,\mathcal{D}}\Bigl[
   -C_{10}(g_N-\chi_N)^2 - C_{01}\,g_N\,\De
   \\&+ C_{10}\,\De(\Dc+\De)\Bigr], \label{eq:C20}\\[2pt]
C_{02} =& \frac{D_c\,(g_N-\chi_N)}{\rt\,\mathcal{D}}\Bigl(
   C_{10}\,g_N - C_{01}\,\Dc \Bigr), \label{eq:C02}\\[2pt]
C_{11} =& \frac{D_c\,\De}{\mathcal{D}}\Bigl(
   -C_{10}\,g_N + C_{01}\,\Dc \Bigr)\,, \label{eq:C11}
\end{align}
\end{subequations}
where
\begin{equation}
\label{eq:DenomCD}
\mathcal{D} \equiv -2 g_N \chi_N \Dc + g_N^2(\Dc+\De)
+ \Dc\bigl(\chi_N^2 - \De(\Dc+\De)\bigr).
\end{equation}

Because the full formulas for $T$ and $g^{(2)}(0)$ followed by Eq.~\eqref{eq:T_general} and Eq.~\eqref{eq:g2_general} are algebraically lengthy, we display the analytic curves as dashed orange lines in the figures. The explicit closed forms are given in the following case for completeness.

\subsubsection{Nonlinear coupling (\(\alpha=2\))}
 As discussed in Sec.~\ref{sec:NHmodel}, the effective low-excitation Hilbert space includes only the states \(\ket{00}\), \(\ket{10}\), \(\ket{20}\), and \(\ket{01}\), and the corresponding ansatz is given in Eq.~\eqref{eq:ansatz_general}. Inserting this ansatz into the non-Hermitian Schrödinger equation with cavity drive Eq.~\eqref{eq:ME_NH}, we obtain the steady-state equations
\begin{subequations}\label{eq:cd_nl_eom}
\begin{align}
i\,\dot{C}_{00} &= D_c\, C_{10}, \\[2pt]
i\,\dot{C}_{10} &= \tilde{\Delta}_c\, C_{10} + D_c\, C_{00} + \sqrt{2}\, D_c\, C_{20}, \\[2pt]
i\,\dot{C}_{20} &= 2\tilde{\Delta}_c\, C_{20} + \sqrt{2}\, D_c\, C_{10}
                 + \sqrt{2}\, g_N\, C_{01}, \\[2pt]
i\,\dot{C}_{01} &= \tilde{\Delta}_e\, C_{01} + \sqrt{2}\, g_N\, C_{20}.
\end{align}
\end{subequations}
Under weak driving conditions \((|C_{00}| \gg |C_{10}| \gg |C_{20}|, |C_{01}|)\), the contribution of the two-excitation amplitudes to \(C_{20},C_{01}\) is negligible. Setting \(\dot{C}_i = 0\), we solve for the steady-state amplitudes in terms of \(C_{00}\),

\begin{subequations}\label{eq:cd_nl_ss}
\begin{align}
C_{10} &\approx -\,\frac{D_c}{\tilde{\Delta}_c}\, C_{00}, \label{eq:cd_nl_C10}\\[4pt]
C_{20} &\approx \frac{\sqrt{2}\,\tilde{\Delta}_e\, D_c}{\,2\,(g_N^2-\tilde{\Delta}_c \tilde{\Delta}_e)\,}\, C_{10},
\label{eq:cd_nl_C20}\\[4pt]
C_{01} &\approx -\,\frac{D_c\, g_N}{\,g_N^2-\tilde{\Delta}_c \tilde{\Delta}_e\,}\, C_{10}.
\label{eq:cd_nl_C01}
\end{align}
\end{subequations}

Substituting the solutions above into Eq.~\eqref{eq:T_general} and Eq.~\eqref{eq:g2_general} , we obtain analytical expressions for the transmission and second-order correlation function:
\begin{align}
T &= \left( \frac{D_c^2 \tilde{\Delta}_e}{(g_N^2 - \tilde{\Delta}_c \tilde{\Delta}_e)^2} + 1 \right) \frac{\gamma_{\mathrm{c}}\gamma_e}{\tilde{\Delta}_c}, \label{eq:cd_nl_T}\\[4pt]
g^{(2)}(0) &= \left( \frac{ \tilde{\Delta}_c \tilde{\Delta}_e }{ (g_N^2 - \tilde{\Delta}_c \tilde{\Delta}_e) \left( 1 + \dfrac{D_c^2 \tilde{\Delta}_e^2}{(g_N^2 - \tilde{\Delta}_c \tilde{\Delta}_e)^2} \right) } \right)^2. \label{eq:cd_nl_g2}
\end{align}
Using the complex detunings \(\tilde{\Delta}_{c(e)} = -i\gamma_{c(e)}/2\), for both channal losses \(\gamma_c = 2\gamma_e = 2\gamma\). We find the spectral shape from T is primarily determined by the denominator term \((\Delta_c - i\gamma)^2\), leading to a resonance peak centered at \(\Delta_c = 0\) with linewidth set by the dissipation rate \(\gamma\). This structure closely resembles a Lorentzian lineshape, consistent with the feature observed in Fig.~\ref{fig:HP&Nan_C_drive}. The closed-form expression in this limit reads
\begin{equation}
    T = \frac{\gamma^2 \left(1 - \dfrac{D_c^2\,(\gamma + 2i\Delta_e)^2}{\left[2g_N^2 + (\gamma + i\Delta_c)(\gamma + 2i\Delta_e)\right]^2} \right)}{(\Delta_c - i\gamma)^2}.
\end{equation}

To gain further insight, we evaluate the correlation at resonance, \(\Delta_c = \Delta_e = 0\), we find
\begin{equation}
g^{(2)}(0)\big|_{\Delta_{c,e}=0} = \left[\frac{\gamma^2 \left(\gamma^2 + 2g_N^2\right)}{ \left(\gamma^2 + 2g_N^2\right)^2 - \gamma^2 D_c^2 } \right]^2.
\label{eq:cd_nl_g2_res}
\end{equation}
In the large-\(N\) limit, where the collective coupling strength \(g_N = g \sqrt{N} \to \infty\), the expression simplifies to
\begin{equation}
\lim_{N \to \infty} g^{(2)}(0) = \frac{\gamma^4/4}{g^4 N^2} \propto \frac{1}{N^2}.
\label{eq:cd_nl_g2_scaling}
\end{equation}
The scaling is independent of \(D_c\), which becomes negligible compared to the collective coupling \(g_N\) in this regime.

To validate the analytical results, we numerically solve the master equation with the HP Hamiltonian Eq.~\eqref{eq:H_HPcd} using QuTiP. The parameters used in the simulation are listed in the Fig.\ref{fig:HP&Nan_C_drive} captions. We also simulate the Full quantum two-photon coupling model and find excellent agreement with the analytical curves in the weak-driving regime in Fig.~\ref{fig3:FQ vs NH}. Deviations at stronger drive strengths arise due to the breakdown of the low-excitation approximation.


\subsection{Emitter Drive}
\subsubsection{Linear coupling (\(\alpha=1\))}
We now consider the emitter driven case under One-photon linear coupling (\( \alpha = 1 \)). The analytical solution proceeds as in the cavity-drive case: we insert the low-excitation ansatz Eq.~\eqref{eq:ansatz_general}  into the time-independent Schrödinger equation governed by the Emitter-drive non-Hermitian Hamiltonian Eq.~\eqref{eq:ME_NH}. The corresponding equations of motion follow from projecting onto the truncated Hilbert space. We obtain the following coupled equations for the amplitudes \( C_{ij} \):
\begin{subequations}
\label{eq:qd_lin_eom}
\begin{align}
i \dot{C}_{00} &= D_e\, C_{01}, \label{eq:qd_lin_eom_a}\\
i \dot{C}_{10} &= \tilde{\Delta}_c\, C_{10} + g_N\, C_{01} + D_e\, C_{11}, \label{eq:qd_lin_eom_b}\\
i \dot{C}_{01} &= \tilde{\Delta}_e\, C_{01} + g_N\, C_{10} + D_e\, C_{00} + \sqrt{2}\, D_e\, C_{02}, \label{eq:qd_lin_eom_c}\\
i \dot{C}_{20} &= 2\tilde{\Delta}_c\, C_{20} + \sqrt{2}\, g_N\, C_{11}, \label{eq:qd_lin_eom_d}\\
i \dot{C}_{02} &= 2\tilde{\Delta}_e\, C_{02} + \sqrt{2}\, D_e\, C_{01} + \sqrt{2}\, (g_N - \chi_N)\, C_{11}, \label{eq:qd_lin_eom_e}\\
i \dot{C}_{11} &= (\tilde{\Delta}_c + \tilde{\Delta}_e)\, C_{11} + D_e\, C_{10} + \sqrt{2}\, g_N\, C_{20} \\&+ \sqrt{2}\, (g_N - \chi_N)\, C_{02}. \label{eq:qd_lin_eom_f}
\end{align}
\end{subequations}

Using the same weak-driving hierarchy \( |C_{00}| \gg |C_{10}|, |C_{01}| \gg |C_{20}|, |C_{02}|, |C_{11}| \), we neglect the influence of two-excitation amplitudes on the lower manifold. Solving the single-excitation sector yields:
\begin{subequations}
\label{eq:qd_lin_C1}
\begin{align}
C_{01} &\approx -\frac{g_N}{\tilde{\Delta}_c}\, C_{10}, \\
C_{10} &\approx -\,\frac{D_e}{\tilde{\Delta}_e - \dfrac{g_N^2}{\tilde{\Delta}_c}}\, C_{00}.
\end{align}
\end{subequations}
To simplify the lengthy expressions for the two-excitation amplitudes \( C_{20}, C_{02}, C_{11} \), we define the following composite quantities:
\begin{align}
\mathcal{Q}' &\equiv \gamma_c + \gamma_e + 2i(\Delta_c + \Delta_e), \\
\mathcal{Q} &\equiv \left(4 g_N^2 + (\gamma_c + 2i\Delta_c)(\gamma_e + 2i\Delta_e)\right) \nonumber\\
&\quad  \left[4 g_N^2\, \mathcal{Q}' - 8 g_N \chi_N\, (\gamma_c + 2i\Delta_c) \right. \nonumber\\
&\quad \left. + (\gamma_c + 2i\Delta_c)\left(4 \chi_N^2 + (\gamma_e + 2i\Delta_e)\, \mathcal{Q}'\right)\right].
\end{align}

Using these, we can express the two-excitation coefficients as:
\begin{subequations}\label{eq:qd_lin_2exc}
\begin{align}
C_{20} &= \frac{8\sqrt{2} D_e^2\, g_N\, \left[g_N\, \mathcal{Q}' - \chi_N\, (\gamma_c + 2i\Delta_c) \right]}{\mathcal{Q}}, \\[4pt]
C_{02} &= -\frac{2\sqrt{2} D_e^2\, (\gamma_c + 2i\Delta_c)\left[4 g_N \chi_N + (\gamma_c + 2i\Delta_c)\, \mathcal{Q}' \right]}{\mathcal{Q}}, \\[4pt]
C_{11} &= \frac{8i D_e^2\, (\gamma_c + 2i\Delta_c)\left[g_N\, \mathcal{Q}' - \chi_N\, (\gamma_c + 2i\Delta_c) \right]}{\mathcal{Q}}.
\end{align}
\end{subequations}
Because the full formulas for $T$ and $g^{(2)}(0)$ followed by Eq.~\eqref{eq:T_general} and Eq.~\eqref{eq:g2_general} are algebraically lengthy, we display the analytic curves as dashed orange lines in the figures. The explicit closed forms are given in the following case for completeness.

\subsubsection{Nonlinear coupling (\(\alpha=2\))}
We now consider the Emitter-driven case under two-photon nonlinear coupling (\( \alpha = 2 \)). We assume the low-excitation relevant ansatz Eq.~\eqref{eq:ansatz_general} and solve the time-independent Schrödinger equation with the non-Hermitian Hamiltonian Eq.~\eqref{eq:ME_NH}.  
Following the same procedure, we obtain the coupled equations governing the amplitudes of the low-excitation states:
\begin{subequations}\label{eq:qd_nl_eom}
\begin{align}
i \dot{C}_{00} &= D_e\, C_{01}, \\
i \dot{C}_{10} &= \tilde{\Delta}_c\, C_{10}, \\
i \dot{C}_{20} &= 2\tilde{\Delta}_c\, C_{20} + \sqrt{2}\, g_N\, C_{01}, \\
i \dot{C}_{01} &= \tilde{\Delta}_e\, C_{01} + \sqrt{2}\, g_N\, C_{20} + D_e\, C_{00}.
\end{align}
\end{subequations}
In steady state (\( \dot{C}_i = 0 \)), one finds \( C_{10} = 0 \). This shows that, in contrast to previous cases, here quantum jumps cannot be neglected even in the weak-drive regime, as they qualitatively affect the steady-state photon number. A detailed discussion of this effect is presented in Appendix~\ref{app:B}.
This issue can be solved by introducing the norm constraint:
\begin{equation}
|C_{10}|^2 + |C_{20}|^2 + |C_{01}|^2 \approx 1 - |C_{00}|^2 \sim D_e^2.
\end{equation}
The analytical solutions are:
\begin{subequations}\label{eq:qd_nl_amps}
\begin{align}
C_{10} &= \frac{D_e \sqrt{ \tilde{\Delta}_c^2 (\tilde{\Delta}_e^2 - 1) + g_N^4 - \frac{1}{2} g_N^2 (4 \tilde{\Delta}_c \tilde{\Delta}_e + 1) }}{g_N^2 - \tilde{\Delta}_c \tilde{\Delta}_e}, \\
C_{01} &= \frac{D_e\, \tilde{\Delta}_c}{g_N^2 - \tilde{\Delta}_c \tilde{\Delta}_e}, \\
C_{20} &= -\frac{D_e\, g_N}{\sqrt{2}(g_N^2 - \tilde{\Delta}_c \tilde{\Delta}_e)}.
\end{align}
\end{subequations}
We then find:
\begin{equation}
T = \frac{\gamma_c \gamma_e \langle a^\dagger a \rangle}{\alpha D_e^2}
= \frac{\gamma^2}{ D_e^2} \left( |C_{10}|^2 + 2 |C_{20}|^2 \right),
\end{equation}
where we have set \( \gamma_c =2 \gamma_e = 2\gamma \), and used \( \alpha = 2 \) in the normalization of the transmission rate.

Due to the algebraic complexity of the full expressions, we report only the resonance case \( \Delta_c = \Delta_e = 0 \). In this limit, the expressions simplify and reveal key physical features. The transmission becomes:
\begin{equation}
T \big|_{\Delta_{c,e}=0}  = 
\frac{
2D_q^2 \left[ 4\, g_N^2 + 4\, \mathcal{A}^{3/4} \right]
}{
4 g_N^4 + 4 g_N^2 \gamma^2 + \gamma^4
},
\label{eq:T_resonance_qubitdrive_nl_correct}
\end{equation}

At resonance, the correlation function becomes:
\begin{equation}
g^{(2)}(0) \big|_{\Delta_{c,e}=0}  = 
\frac{256(4g_N^5+4g_N^3 \gamma^2+g_N \gamma^4)^2} {D_e^2
 \mathcal{W}
 \left[ 16 g_N^2 + 16\, \mathcal{A}^{3/4} \cdot \mathcal{R} \right]^2
},
\label{eq:g2_resonance_qubitdrive_nl}
\end{equation}

where
\begin{align}
\mathcal{A} &\equiv g_N^4 + \frac{1}{2} g_N^2 (2\gamma^2 - 1) + \frac{1}{4} \gamma^2 (4 + \gamma^2), \\
\mathcal{R} &\equiv \cos\left( \frac{1}{2} \arg(\mathcal{A}) \right) - i \sin\left( \frac{1}{2} \arg(\mathcal{A}) \right).\\
\mathcal{W}&= g_N^2 \left(16 g_N^4 + 8 g_N^2 \gamma^2 + \gamma^4\right)
\end{align}
To analyze two-photon blockade, we compute the third-order correlation function, defined accordingly to Eq.~\eqref{eq: correlation function}, by extending the wavefunction ansatz to include higher-excitation states.
\begin{equation}
\begin{split}
\label{ansatz_3ph}
    |\psi(t)\rangle = &C_{00}(t) |00\rangle + C_{10}(t) |10\rangle + C_{20}(t) |20\rangle \\&+ C_{30}(t) |30\rangle+C_{01}(t) |01\rangle+C_{11}(t) |11\rangle.
\end{split}
\end{equation}
We find for the third-order correlation function,
\begin{equation}
g^{(3)}(0) = \frac{6 |C_{30}|^2}{\left( |C_{10}|^2 + 2 |C_{20}|^2 + 3 |C_{30}|^2\right)^3}.
\end{equation}
The numerically computed values at resonance, for fixed emitter number and drive strength, are shown in Fig.~\ref{fig:g2g3_HP}.

\section{Validity of non-Hermitian approximation}
\label{app:B}

In the main text, we analyze the system dynamics using an effective non-Hermitian Hamiltonian, which neglects quantum jump terms \( \mathcal{J}[L](\rho) = L \rho L^\dagger \) of the Lindblad master equation. This approach is commonly justified in the weak-driving regime. However, its validity depends not only on the drive strength but also on the system’s level structure and whether dissipation contributes to populating relevant states.
In the case of cavity driving, this approximation remains valid across all the scenarios discussed. However, when the emitters are driven, the validity of this approximation depends on the type of light–matter coupling. In this appendix, we analyze and compare two such cases — linear coupling (\( \alpha = 1 \)) and nonlinear two-photon coupling (\( \alpha = 2 \)) — and identify the conditions under which the non-Hermitian approximation fails. In general, quantum jump terms can be neglected if they only induce higher-order corrections to state populations or observables.  This is the case for the linear coupling model (\( \alpha = 1 \)), where all excited states relevant to observables like photon number \( \langle a^\dagger a \rangle \) are already populated via coherent dynamics. For $\alpha = 2$ instead, the state  \( |10\rangle \) is populated only as a result of decay process from the state \( |20\rangle \). Let us discuss the two cases separately.
\paragraph*{Linear Coupling Case (\( \alpha = 1 \))} --
Under weak emitter drive \( D_e \ll 1 \), and linear coupling \( g_N (a^\dagger b + a b^\dagger) \), the Hamiltonian generates the following transitions:
\[
|00\rangle \xrightarrow{D_e} |01\rangle \xrightarrow{g_N} |10\rangle \xrightarrow{D_e} |11\rangle \xrightarrow{g_N} |20\rangle.
\]
All states contributing to \( \langle a^\dagger a \rangle = |C_{10}|^2 + 2|C_{20}|^2 \) are populated coherently. Quantum jumps such as:
\[
\mathcal{J}[a](\rho): \quad |20\rangle \to |10\rangle, \quad |10\rangle \to |00\rangle,
\]
modify the amplitudes only at higher orders. The non-Hermitian approximation is therefore valid.\\

\paragraph*{Nonlinear Coupling Case (\( \alpha = 2 \))} --
In the two-photon coupling case, with interaction \( g_N(a^{\dagger 2} b + a^2 b^\dagger) \), the Hamiltonian induces the following transitions:
\[
|00\rangle \xrightarrow{D_e} |01\rangle \xrightarrow{g_N} |20\rangle.
\]
However, no coherent process exists that connects to the state \( |10\rangle \). Thus, \( C_{10} = 0 \) under purely non-Hermitian Hamiltonian dynamics. Nevertheless, full numerical simulations confirm a finite \( C_{10} \), which arises due to the quantum jump:
\[
\mathcal{J}[a](\rho): \quad |20\rangle \to |10\rangle.
\]
This process occurs in the same order as the coherently populated states \( |01\rangle \) and \( |20\rangle \), all scaling as \( \mathcal{O}(D_e^2) \). However, leakage to the  \( |01\rangle \) state can be effectively introduced by introducing a constraint on the normalization. 
Under weak-drive regime the system predominantly occupies the ground state \( |00\rangle \), and excited state population are of the order of
\begin{equation}
|C_{10}|^2 + |C_{20}|^2 + |C_{01}|^2 \approx \epsilon.
\end{equation}
where we defined
\begin{equation}
\epsilon = 1 - |C_{00}|^2 \approx \sum_i |C_i|^2 \sim D_{c/e}^2.
\end{equation}
This constraints allows to correctly account for the population in the \( |01\rangle \) state, as confirmed by numerical simulations of the HP model.


\clearpage
\newpage
\bibliography{References}

\end{document}